\documentclass[article]{IEEEtran}
\usepackage{xspace,empheq,fancybox,amsmath,bbm,amssymb,epsfig,subfigure,syntonly,times,amsthm} \usepackage{psfrag,color,bm,array,cite}
\usepackage{url,cite,footnote,xspace,syntonly,algorithm,algpseudocode}
\usepackage{verbatim,multirow,slashbox}
\usepackage[T1]{fontenc}
\usepackage{hyperref}
\input{mysymbol.sty}

\definecolor{darkblue}{rgb}{0.0,0.0,0.6}
\hypersetup{colorlinks,breaklinks,
            linkcolor=darkblue,urlcolor=darkblue,
            anchorcolor=darkblue,citecolor=darkblue}

\newtheorem{remark}{\bfseries Remark}

\newcommand{\tr}{\mathrm{Tr}}

\newcommand{\diag}{\mathrm{diag}}

\newcommand{\EE}{\mathbb{E}}

\newcommand{\tProj}{\widetilde{\mathbb{P}}}
\newcommand{\tdnabla}{\widetilde{\nabla}}

\newcommand{\nhD}{\bbD^{-\frac{1}{2}}}

\newcommand{\ml}{{}}

\newcommand{\re}{{}}
\newcommand{\T}{\top}
\newcommand{\R}{\mathbb{R}}
\newcommand{\lmax}[1]{\lambda_{\max}\{#1\}}
\newcommand{\lmin}[1]{\lambda_{\min}\{#1\}}

\begin{document}
\title{Decentralized Dynamic Optimization\\for Power Network Voltage Control}
\author{Hao Jan Liu, Wei Shi, and Hao Zhu\thanks{\protect\rule{0pt}{3mm}
This material is based upon work supported by the Department of Energy under Award Number DE-OE000078 and the National Science Foundation under Award Number EPCN-1610732. The authors are with the University of Illinois at Urbana-Champaign, 306 N. Wright Street, Urbana, IL, 61801, USA; Emails: {\{haoliu6,wilburs,haozhu\}{@}illinois.edu}.
}}
\renewcommand{\thepage}{}
\maketitle
\pagenumbering{arabic}
\begin{abstract}
Voltage control in power distribution networks has been greatly challenged by the increasing penetration of volatile and intermittent devices. These devices can also provide limited reactive power resources that can be used to regulate the network-wide voltage. A decentralized voltage control strategy can be designed by minimizing a quadratic voltage mismatch error objective using  gradient-projection (GP) updates. Coupled with the power network flow, the local voltage can provide the instantaneous gradient information. This paper aims to analyze the performance of this decentralized GP-based voltage control design under two dynamic scenarios: i) the nodes perform the decentralized update in an asynchronous fashion, and ii) the network operating condition is time-varying. For the asynchronous voltage control, we improve the existing convergence condition by recognizing  that the voltage based gradient is always up-to-date. By modeling the network dynamics using an autoregressive process and {\ml considering time-varying resource constraints}, we provide an error bound in tracking the instantaneous optimal solution to the quadratic error objective. This result can be extended to more general \textit{constrained dynamic optimization} problems with smooth strongly convex objective functions under stochastic processes that have bounded iterative changes. Extensive numerical tests have been performed to demonstrate and validate our analytical results for realistic power networks.
\end{abstract}

\section{Introduction}\label{sec:intro}
The smart grid vision has led to continued proliferation of distributed energy resources (DERs) in the power distribution networks, including the rooftop photovoltaic (PV) panels and batteries of electric vehicles. Albeit environmentally friendly, the DERs could greatly challenge the operational goal of maintaining a satisfactory voltage level per power system reliability standards. High variability of PV generation and elastic loads can cause unexpected network-wide voltage fluctuations, at a much faster dynamics than the time-scale of traditional voltage regulation devices; see e.g., \cite{Car_tps08,Tur_proc11,Rob_tps13} and references therein. Thanks to  advances in power electronics,  DER devices can also be excellent sources of \textit{reactive power}, a quantity that is known to have a significant impact on the network voltage level. Thus, one major task for operating future distribution networks is to design effective voltage control strategies to utilize the reactive power from DERs, at minimal hardware requirements.

By and large, the communication infrastructure that supports power distribution networks is, and will continue to be lacking in the foreseen future. Meanwhile, cost concerns for DER products inevitably limit their sensing/computation capabilities. Therefore, voltage control approaches by minimizing a centralized voltage mismatch error are not practically attractive, because their success strongly relies on high-quality communications either between a control center and remote devices \cite{Far_pesgm12}, or among neighboring devices \cite{Rob_tps13,dall_tsg13,lanl}. These optimization-based open-loop control designs may become unstable under communication delays or noises during online implementations. Also, it is unclear whether these approaches will be robust to asynchronous computational speeds among the networked DERs of heterogeneous hardware capabilities. To cope with the limited cyber infrastructure, a decentralized voltage control framework is more preferred for distribution network operations; see e.g., \cite{Car_tps08,Tur_proc11,Far_cdc13,hzml_tps15}. Under this framework, each node only needs to measure its locally available voltage level as the controller input. Our earlier work \cite{hzml_tps15} has offered an overarching framework that generalizes a variety of decentralized control designs, along with convergence analysis for static system scenarios. Interestingly, the local voltage measurement naturally provides the instantaneous gradient information for a centralized error objective by weighting the voltage mismatch. Hence, the decentralized voltage control approach using this measurement boils down to the classical gradient-projection (GP) method accounting for the limits on reactive power resources. This voltage-measurement based decentralized design does not require any real-time communications, and can be implemented with minimal upgrades in sensing hardware.

The goal of this paper is to analyze the performance of this decentralized GP-based voltage control design under two dynamic scenarios: i) the nodes perform the decentralized update in an asynchronous fashion; and ii) the network operating condition is dynamically changing. The scenario of asynchronous updates arises from heterogeneous hardware capabilities among different DERs. It is also motivated to allow the ``plug-and-play'' functionality for flexible DER integration to distribution networks. The classical asynchronous optimization framework always  accounts for the case of outdated information from other nodes, and thus the choice of step-size has to be more conservative due to the information delay; see e.g., \cite{Bertsekas_PD} and \cite{fey_cdc14}. Different from this, the voltage measurement for our decentralized control design always provides the up-to-date gradient information and does not suffer any information delay.  Thanks to this physical power network coupling, we can show that the choice of the step-size for the asynchronous decentralized updates is the same to the synchronous case. Hence, its convergence condition is robust to potential discrepancy in the control update  rates among different DERs.

{\ml In power networks, voltage control designs under time-varying operating conditions have been implemented as the static optimal power flow solutions to dynamic settings in a heuristic fashion \cite{Scott_14,Gill_tps14}. The scope of these efforts is more centered around dynamic voltage control implementations rather than providing control performance guarantees. The latter is of high interest when accounting for the variability of networked generations and loads in practice.} With a time-varying objective function,  this problem becomes a \textit{stochastic optimization} one; see e.g., \cite{sla_spm14}. Stochastic approximation algorithms such as stochastic (sub-)gradient decent have been developed in e.g., \cite{bottou_ccs2010,Johnson_NIPS13} and have been adopted by \cite{vkgg_tps15} for this voltage control problem. Nonetheless, performance analysis for stochastic optimization algorithms has focused on the convergence to the optimal solution that minimizes the expected objective function \cite{sla_spm14}. Aiming at the error bound in tracking the instantaneous optimal solution, our analysis is more closely related to the body of work on \textit{dynamic convex optimization}; see e.g., \cite{Ling_tsp14,tow_tsp14,sim_arxiv15}. This type of problems  typically arises from applications in autonomous teams and wireless sensor networks, such as target tracking \cite{Ke_tr11} and estimation of the stochastic path \cite{Jaku_tsp13}. Some of these dynamic optimization algorithms follow a gradient descent update, but none of them has considered the formulation of constrained  optimization. This is the key difference from our voltage control problem since the control input has to be feasible under the {\ml dynamic} reactive power limits. Hence, the main contribution of our work lies in the fact that it explicitly accounts for the {\ml time-varying} projection operation of \textit{constrained} dynamic optimization. Our tracking error performance bounds will be derived for a  quadratic objective function and an autoregressive dynamic model, motivated by this specific voltage control problem. Nonetheless, our analytical results can be extended to more general constrained dynamic optimization problems with smooth strongly convex objective functions under stochastic processes that have bounded iterative changes.

The remainder of this paper is organized as follows. Section \ref{sec:model} presents the modeling of power distribution networks as the basis of our analysis, while Section \ref{sec:stochvar} designs a decentralized voltage control strategy using the GP updates.  Performance analysis under  asynchronous updates  or time-varying network operating condition is offered in Section \ref{sec:asyn_intro} and Section \ref{sec:conana}, respectively. Section \ref{sec:num} presents numerical results to demonstrate and validate our analytical results, and the paper is wrapped up in Section \ref{sec:con}.

\section{System Modeling} \label{sec:model}
\begin{figure}[tb]
	\centering
	\vspace{3pt}
	\includegraphics[width=0.9\linewidth,clip = true, trim = .8in 2.88in 1.5in 2.36in]{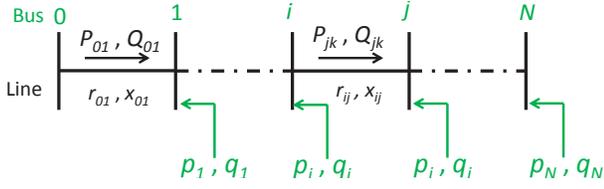}
	\caption{A radial distribution network with bus and line associated variables.}
	\label{fig:radial}
	\vspace{-4mm}
\end{figure}
A power distribution network can be modeled using a tree-topology graph $(\ccalN,\ccalE)$ with the set of buses (nodes) $\ccalN:=\{0,...,N\}$ and the set of line segments (edges) $\ccalE:=\{(i,j)\}$; see Fig. \ref{fig:radial} for a radial network illustration. At every bus $j$, let $v_{j}$ denote its voltage magnitude and $p_j$ $(q_j)$ represent the real (reactive) power injection, respectively. Bus $0$ corresponds to the point of common coupling,  assumed to  have unity reference voltage; i.e., $v_0=1$. For each line $(i,j)$, let $r_{ij}$ and $x_{ij}$ denote its resistance and reactance and $P_{ij}$ and $Q_{ij}$ the power flow from $i$ to $j$, respectively.

To tackle the nonlinearity of power flow models, one can assume negligible line losses and almost flat voltage, i.e., $v_j \cong 1,~\forall j$. Under these assumptions, the so-termed LinDistFlow model has been developed in \cite{Baran_tpd89}, and its accuracy can be numerically corroborated by several recent work \cite{Far_cdc13, Sulc_tec14, Li_tse15 ,hzml_tps15,Robbins_tps16}. For each $(i,j)$, the LinDistFlow model asserts the bus power balance and line voltage drop, as given by
\begin{subequations}
	\begin{align}
P_{ij}-\sum_{k \in \mathcal{N}_{j}^{+}}P_{jk} &= -p_j, \label{cp} \\
Q_{ij}-\sum_{k \in \mathcal{N}_{j}^{+}}Q_{jk} &= -q_j,\label{cq} \\
	v_{i}-v_{j} &= r_{ij}P_{ij}+x_{ij}Q_{ij} \label{cv}
	\end{align}
\label{lindist}
\end{subequations}
where the neighboring bus set $\mathcal{N}_{j}^{+}:=\{k | (j,k)\in \mathcal{E}$, and $k$ is downstream from $j\}$. For example, we have $\ccalN_i^+=\{j\}$ in Fig. \ref{fig:radial}.

To construct the matrix form of \eqref{lindist}, denote the graph incidence matrix using the $(N+1)\times N$ matrix $\bbM^o$. Each one of its $\ell$-th column corresponds to a line $(i,j)$, with all zero entries except for the $i$-th and $j$-th (see e.g., \cite[pg. 6]{west_book}). 
Let us set $M^o_{i\ell} =1$ and $M^o_{j\ell} = -1$ if $j\in\ccalN_i^+$. Let $\bbm_0^\T$ represent the first row of $\bbM^o$ corresponding to bus 0, with the rest of the rows in the $N\times N$ submatrix $\bbM$. Accordingly, $\bbM$ is full-rank and invertible because the network is connected under the tree-topology assumption \cite{west_book}. Upon concatenating all scalar variables into vector form, one can represent \eqref{lindist} as
\begin{subequations}
\label{ldf}
	\begin{align}
	-\bbM \bbP &= -\bbp, \label{ldfPm} \\
	-\bbM \bbQ &= -\bbq, \label{ldfQm} \\
	\bbm_0 + \bbM^\T \bbv &=  \bbD_r \bbP + \bbD_x \bbQ \label{ldfVm}
	\end{align}
\end{subequations}
where the $N\times N$ diagonal matrix $\bbD_r$ has diagonals equal to all line $r_{ij}$'s; and similarly for $\bbD_x$ having all $x_{ij}$'s. Viewing the uncontrollable $\bbp$ as a constant, one can solve for $\bbP$ and $\bbQ$ in \eqref{ldf} to establish the following:
\begin{align}
\bbv = \bbX\bbq + \bbarbbv \label{ldfVm2}
\end{align}
where the nominal voltage vector $\overline{\bbv}$ captures the effects of $\bbp$ when  $\bbq = \bb0$, while $\bbX := (\bbM^\T)^{-1}\bbD_x \bbM^{-1}$ can be viewed as the network reactance matrix. The linear model \eqref{ldfVm2} constitutes the basis for developing decentralized voltage control algorithms.

\begin{remark}[Modeling Considerations] Nonlinearity of the power flow model could be tackled by the formulation of semidefinite programming (SDP) \cite{brhzadg_tps15}. Generally, a rank relaxation approach is adopted in order to obtain a SDP convex problem formulation. Nonetheless, the SDP-based power flow formulation would face several challenges when applied to power (distribution) networks in practice. First, the resultant power flow solution could be non-exact \cite{edhzgg_tsg13} under this setting when power networks are unbalanced multi-phase. Furthermore, performance guarantees for the SDP approach fail to hold for general network topology such as meshed systems \cite{Low_tcns14}. Last, the SDP solution increases the computational burden significantly as the size of the network grows. Different from the aforementioned modeling approach, the linearized model \eqref{ldfVm2} holds for more realistic scenarios including meshed topology and unbalanced three-phase networks \cite{hzml_tps15,kekatos_tps16} at minimal computational complexity requirements. Hence, the ensuing analysis can be readily generalized to practical power networks, as verified by numerical tests later on in Sec. \ref{sec:num}.
\end{remark}

\section{Decentralized Voltage Control} \label{sec:stochvar}

The goal is to control the reactive power $\bbq$, such that $\bbv$ approaches a given desirable voltage profile $\bbmu$. The flat voltage profile is typically chosen; i.e., $\bbmu := \mathbf{1}$. At every time instance $k$, let $\bbarbbv_k$ denote the instantaneous nominal voltage profile. To allow for a decentralized control design, it turns out that one can minimize a weighted voltage mismatch error using $\bbB:=\bbX^{-1}$, as given by
\begin{align}\label{wvobj}
\bbq_{k}^* = \arg\min_{\bbq \in \ccalQ_k} f_k(\bbq) := \frac{1}{2} (\bbX \bbq + \bbarbbv_k - \bbmu)^\T \bbB (\bbX \bbq + \bbarbbv_k - \bbmu)
\end{align}
where the constraint set $\ccalQ_k := \{\bbq\big|\bbq \in [\underline{\bbq}_k,\overline{\bbq}_k] \}$ accounts for the time-varying limits of local reactive power resources at every bus \cite{Tur_proc11}. Interestingly, matrix $\bbB = \bbM \bbD_x^{-1} \bbM^\T$, by definition, is {\ml a weighted, reduced} graph Laplacian for $(\ccalN,\ccalE)$. Since all the reactance values are positive, $\bbB$ is symmetric and positive definite \cite{hzml_tps15}. Accordingly, the weighted voltage mismatch error objective of \eqref{wvobj} is convex, and in fact quadratic, in the variable $\bbq$. Ideally, the unweighted error norm $\|\bbv-\bbmu\|$ is the best objective in order to achieve the flat voltage profile. Furthermore, compared to the traditional paradigm of maintaining the voltage within limits, this unweighted objective can improve the system-wide voltage profile by coordinating network-wide VAR resources. This is attractive for energy saving programs such as a conservation voltage reduction implementation \cite{mlrm_sgcomm14}. Albeit the problem \eqref{wvobj} minimizes a surrogate objective, it has been shown in \cite{hzml_tps15} that $\bbq^*_k$ can closely approximate the optimal solution to the ideal unweighted error norm, especially if  there are abundant reactive power resources.

Thanks to the separable structure of the box constraint $\ccalQ_k$, the gradient-projection (GP) method \cite[Sec. 2.3]{Ber_NPbook} can be invoked to solve  \eqref{wvobj}.
Upon forming its instantaneous gradient $\nabla f_k(\bbq_k) := \bbX \bbq_k + \bbarbbv_k - \bbmu$, the GP iteration for a given positive step-size $\epsilon>0$ becomes
\begin{align}
\bbq_{k+1} = \mathbb P_k \left[\bbq_k - \epsilon \bbD \nabla f_k(\bbq_k) \right]
\label{graprojs}
\end{align}
where the projection operator $\mathbb P_k$ thresholds any input to be within $\ccalQ_k$, and $\bbD$ is a diagonal scaling matrix that can be designed. As a first-order method, the GP method has linear convergence rate, while the convergence speed depends on the condition number of the corresponding Hessian matrix \cite[Sec. 3 .3]{Ber_NPbook}. Motivated by this fact, the scaling matrix can be chosen according to the inverse of the diagonals of Hessian matrix by setting $\bbD :=[\diag(\bbX )]^{-1}$ to approximate the Newton gradient. Note that a positive diagonal matrix $\bbD$  affects neither the separability of operator $\mathbb P$, nor the optimality of the update \eqref{graprojs}.

By setting the GP iterate $\bbq_k \in \ccalQ_k$ to be the control input at any time $k$, the instantaneous voltage becomes
$\bbv_k = \bbX \bbq_k+\bbarbbv_k$ based on \eqref{ldfVm2}. Thanks to the physical power network coupling, $\bbv_k$ always provides the up-to-date gradient information as $\nabla f_k(\bbq_k) = \bbv_k-\bbmu$. Accordingly, the GP update in \eqref{graprojs} can be implemented by directly measuring the instantaneous voltage as
\begin{align}\label{graprojsD}
\bbq_{k+1} = \mathbb P_k \left[\bbq_k - \epsilon \bbD (\bbv_k - \bbmu) \right],
\end{align}
which can be completely decoupled into decentralized updates at each bus because $\mathbb P_k$ is separable. This decentralized voltage control is very attractive with minimal hardware requirements as each bus only needs to measure its local voltage {\ml and requires no communication.} The optimality and convergence conditions for \eqref{graprojsD} have been investigated in \cite{hzml_tps15}, which are summarized by the following proposition.

\begin{proposition}\label{prop:loc}
When $\bbarbbv_k=\bbarbbv$ and $\ccalQ_k=\ccalQ$ (time-invariant case), the decentralized update \eqref{graprojsD} approaches the unique time-invariant optimizer $\bbq^*$ of problem \eqref{wvobj} if the step-size $\epsilon\in(0,2/M)$ where
\begin{equation}\label{eq:M}
M:=\lmax{\tdbbX}
\end{equation}
is the largest eigenvalue of matrix
\begin{equation}\label{eq:tX}
\tdbbX:=\bbD^{\frac{1}{2}}\bbX\bbD^{\frac{1}{2}}.
\end{equation}
\end{proposition}

The decentralized voltage control design has to account for a variety of uncertainties in practical system implementations. First, due to heterogeneity of various DERs, it is difficult to perfectly synchronize the decentralized update at different buses. This is especially important to facilitate  the ``plug-and-play'' functionality  for flexible DER integration. Second, the volatility and intermittence of electric loads and renewable-based generations challenge the static setting of time-invariant $\bbarbbv_k$. It is of considerable interest to quantify the performance of the decentralized control \eqref{graprojsD} in terms of tracking the time-varying optimizer $\bbq_k^*$ to the dynamic objective $f_k$.

\section{Asynchronous Decentralized Voltage Control}\label{sec:asyn_intro}
When the DERs are heterogeneous, it is logical that the decentralized voltage update should be performed in an asynchronous fashion. This way, the buses with better computation and sensing capabilities do not need to wait for the slowest one. Accordingly, these buses can execute more updates for a same time interval and hence respond more quickly to localized voltage deviations.

To this end, let the set $\ccalK_j$ collects all the time instances when bus $j$ executes its decentralized update.  The asynchronous counterpart of \eqref{graprojsD} can be modeled by
\begin{align}\label{asgp}
\bbq_{k+1} = \bbq_k + \bbs_k, ~\forall k
\end{align}
with the difference at bus $j$ given by
\begin{align}\label{sjt}
\hspace{-1em}s_{j,k} := \left\{\begin{array}{cl}\!\!\! \mathbb P_{j,k} \left[ q_{j,k} - \epsilon D_j (v_{j,k} - \mu_j) \right] - q_{j,k},& \!k\in \ccalK_j, \\ 0, & \! k\notin \ccalK_j, \end{array} \right.
\end{align}
where $D_j$ is the $(j,j)-$th entry of $\bbD$, while $\mathbb P_{j,k}$ projects the input to $[\underline q_{j,k}, \overline q_{j,k}]$.

To establish the convergence condition, every bus needs to update sufficiently often. Similar to the classical asynchronous algorithm analysis in \cite[Ch. 7]{Bertsekas_PD}, we assume the following \textit{bounded update delay} condition.

\begin{assumption}[Bounded Update Delay]\label{ass:B-Delay}
For every bus $j$ and time instance $k \geq 0$, there exists a positive integer $K$ such that at least one  element  in the set $\{k, k+1,\ldots, k+K-1\}$ belongs to $\ccalK_j$. Equivalently, every bus must update at least once every $K$ iterations.
\end{assumption}

In addition to Assumption \ref{ass:B-Delay}, the analysis of classical asynchronous algorithms also assumes the \textit{bounded information delay} condition \cite[Ch. 7]{Bertsekas_PD}. Due to potential communication delays among peer processors, the updates at some buses may not be executed based on the most up-to-date system-wide information. By assuming that the local information used for computing the gradient  is potentially obsoleted by at most $K$ iterations, it has been established in \cite[Sec. 7.5]{Bertsekas_PD} that the asynchronous GP algorithm in \eqref{asgp}-\eqref{sjt} converges to the optimal solution with more conservative step-size choice given by
\begin{align}
\label{eq:old_step_size}
0<\epsilon< 1/[M(1+K+NK)].
\end{align}
Due to information delay, the choice of step-size would depend on the slowest processor in the network. Generally, this bound on $\epsilon$  can be much smaller than the $2/M$ bound in Proposition \ref{prop:loc}, resulting in a much slower convergence compared with the synchronous case.

For our decentralized voltage control, the gradient $\nabla f_k(\bbq_k) =\bbv_k-\bbmu$ always holds.  Thanks to the physical power network coupling, the local voltage $v_{j,k}$ always provides the up-to-date gradient information at every iteration $k$. Hence, whenever a node is active, the difference $s_{j,k}$ computed in \eqref{sjt} does not suffer from any information delay. This is different from most parallel and distributed algorithms where the updates at every processor require information sent by peer processors. Therefore, convergence of \eqref{asgp}-\eqref{sjt} no longer requires the more conservative  choice of $\epsilon$  in \cite[Sec. 7.5]{Bertsekas_PD}.

\begin{theorem}\label{prop:asyn}
Under Assumption \ref{ass:B-Delay}, when $\bbarbbv_k=\bbarbbv$ and $\ccalQ_k=\ccalQ$ (time-invariant case), the asynchronous update illustrated in \eqref{asgp}-\eqref{sjt} converges to the time-invariant optimizer $\bbq^*$ if the step-size $\epsilon\in(0,2/M)$.
\end{theorem}
\begin{IEEEproof}
First, it is easy to show that the fixed-point of \eqref{asgp}-\eqref{sjt} is the same to \eqref{graprojsD} using contradiction. As for the convergence, by projecting any scalar $q$ to $[\underline q_j, \overline q_j]$, it holds that $[\mathbb P_j (q) - q_{j,k}][\mathbb P_j ( q) - q] \leq 0$ for the iterate $q_{j,k}\in[\underline q_{j,k}, \overline q_{j,k}]$ where $\mathbb P_j=\mathbb P_{j,k}$; see e.g., \cite[Sec. 3.3.1]{Bertsekas_PD}. This implies that at every iteration $k\in\ccalK_j$ (use $f_k=f$ for time-invariant case)
\begin{align}
s_{j,k}[s_{j,k}\! +\! \epsilon D_j \nabla_j f(\bbq_k)]\! = \!\epsilon D_j s_{j,k} \nabla_j f(\bbq_k) + (s_{j,k})^2 \leq 0. \label{projthem}
\end{align}
The descent lemma in \cite[Sec. 3.3.2]{Bertsekas_PD} together with the Lipschitz continuity of $f(\cdot)$ entails  for every $k$
\begin{align*}
f(\bbq_{k+1}) &= f(\bbq_k+\bbs_k)\\
& \leq f(\bbq_k) + \bbs_k^\T \nabla f(\bbq_k) + (M/2) \|\bbs_k\|^2 \\
& \leq f(\bbq_k) - \left(\frac{1}{\epsilon} - \frac{M}{2}\right) \|\bbs_k\|^2. ~~[\mathrm{cf. ~\eqref{projthem}}]
\end{align*}
Summing up the inequality over all iterations yields
\begin{align*}
\textstyle \sum_{k = 0}^\infty \|\bbs_k\|^2 \leq \left(\frac{1}{\epsilon} - \frac{M}{2}\right)^{-1} f(\bbq_0) < \infty,
\end{align*}
which holds as long as  $\left(\frac{1}{\epsilon} - \frac{M}{2}\right)$ is positive. {\ml Thus, if  $0 <\epsilon < 2/M$, $\|\bbs_k\|^2$ is summable and the convergence $\lim_{k\rightarrow \infty} s_{j,k} = 0$ holds for every $j$. And this completes the asymptotic convergence claim for $q_k$ to its fixed point $\bbq^*$.}
\end{IEEEproof}

Convergence analysis for the asynchronous voltage control updates ensures that the heterogeneity of DERs does not affect the choice of $\epsilon$. As for online  implementation, this result can provide guaranteed stability for the proposed control design.

\section{Dynamic Decentralized Voltage Control}\label{sec:conana}
In addition to the asynchronous voltage control updates, the uncertainty in the nominal voltage  $\bbarbbv_k$ further challenges the performance of decentralized voltage control. The volatility and intermittence of loads and generations lead to temporal variations in the network operating condition, i.e., a dynamic $\bbarbbv_k$. Thus, it is imperative to analyze the performance of the decentralized voltage control under a dynamic setting.

To this end, the first order autoregressive (AR(1)) process is adapted to model the short-term dynamics.

\begin{assumption}[Dynamic Voltage Profile]\label{ass:DyVol}
For a given constant vector $\bbarbbc$, the nominal voltage  $\bbarbbv_k$ follows a wide-sense stationary AR(1) process, as given by
\begin{align}
\bbarbbv_{k+1} &= \bbA \bbarbbv_k + \bbeta_{k+1} + \bbarbbc \label{v_ar1}
\end{align}
where $\bbA$ is a time-invariant transition matrix with its spectral radius less than $1$, while $\bbeta_{k+1}$ represents a zero-mean white noise process with covariance matrix $\bbSigma_\eta$.
\end{assumption}

The AR(1) model \eqref{v_ar1} can capture both a short-term temporal and spatial correlation of the nominal voltage profile. Its validity has been corroborated by  \cite{hassan_tps15} from real data based tests. Under Assumption \ref{ass:DyVol}, for every time $k$, the nominal voltage $\bbarbbv_k$ has constant mean $\mathbb{E} \bbarbbv_{k} = (\bbI-\bbA)^{-1}\bbarbbc$  with a bounded covariance matrix $\bbSigma_\bbarv$ satisfying
\[
\bbSigma_\bbarv = \bbA\bbSigma_\bbarv\bbA^\T + \bbSigma_\eta.
\]

{\ml For ease of exposition, the spatial correlation for power networks is not considered as it is often time negligible \cite{cotilla_sj12}, corroborated by the structure of inverse of reduced graph Laplacian matrix $\bbX$. Since $\bbX$ is in fact diagonally dominant, the variations in loads and generations tend to have very localized impacts. However, to be detailed soon, our analysis can be generalized to any higher order AR/stochastic process that has bounded successive differences.} Thus, by ignoring the spatial correlation and assuming uniform variation across buses, the AR(1) model \eqref{v_ar1} can be simplified as follows:
\begin{align}
\bbarbbv_{k+1} &= \alpha \bbarbbv_k + \bbeta_{k+1} + \bbarbbc \label{v_ar2}
\end{align}
with $\bbSigma_\eta = \sigma^2 \bbI$. Accordingly, the stability condition boils down to $|\alpha| < 1$, while $\bbarbbv_{k}$ the mean $\mathbb{E} \bbarbbv_{k} = \bbarbbc / (1-\alpha)$ and the covariance $\bbSigma_\bbarv = \sigma^2/(1-\alpha^2)\bbI$. The smaller the value of $|\alpha|$ is, the faster that the nominal voltage $\bbarbbv_k$ evolves.

{\re \begin{proposition}[Lemma 1 in \cite{liu_ssp16}]\label{lemma:var_dif_v}
Under Assumption \ref{ass:DyVol}, the expectation of the  weighted norm of consecutive difference is bounded, i.e., there exists a bounded constant $B_1$ such that
\begin{equation}\label{eq:var_dif_v}
\EE\|\bbarbbv_{k+1}-\bbarbbv_k\|_{\bbD}^2 =\frac{2\sigma^2\tr\bbD}{1+\alpha} \leq B_1 \text{ for all } k=0,1,\ldots.
\end{equation}
where the weighted norm $\|\bbv\|_\bbD ^2 := \bbv^\T\bbD\bbv$ for any $\bbv$.
\end{proposition}}

{\re Under the settings of both dynamic objective and constraint, we formally state the assumption we make for the performance analysis of the gradient projection approach \eqref{graprojsD}.}

{\re \begin{assumption}[Bounded Drift of Optimizer]\label{ass:B-Drift}
The successive difference of the transient optimal solution is bounded, i.e., there exists some bounded constant $B_2$ such that (see \eqref{wvobj} for the definition of $\bbq_k^*$)
\[
\EE\|\bbq_k^*-\bbq_{k+1}^*\|_{\bbD^{-1}}^2 \leq B_2 \text{ for all } k=0,1,\ldots.
\]
\end{assumption}}
{\re This assumption is related to the boundedness of voltage drift (see Prop. \ref{lemma:var_dif_v}) and the compactness of box constraints. For instance, when the reactive power is unlimited, i.e., $\ccalQ_k=\R^N$, one can easily verify that $\EE\|\bbq_k^*-\bbq_{k+1}^*\|_{\bbD^{-1}}^2$ is bounded. When the reactive power is uniformly limited, i.e., $\ccalQ_k$ is always some compact set (double-sided box constraint suffices) for all $k$, we still have the bounded optimizer drift due to the fact that $\bbq_k^*\in\ccalQ_k$. Albeit the error bound (stability) we are going to construct will depend on $B_2$, intuitively, a smaller voltage drift bound $B_1$ tends to decrease the drift of the optimizer bound $B_2$ in power networks.}

We first introduce a few quantities to simplify the presentation:
\[
\bby_{k}:=\nhD\bbq_{k},\ \bby_{k}^*:=\nhD\bbq_{k}^*,
\]
\[
\bbu_k:=\nhD(\bbarbbv_k-\bbmu),
\]
and $\tProj_k[\cdot]$ is an operator that projects its input onto the set
\[
\widetilde{\ccalQ}_k:=\{\bbq\big|\bbq \in [\bbD^{-\frac{1}{2}}\underline{\bbq}_k,\bbD^{-\frac{1}{2}}\overline \bbq_k]\}.
\]
This way, the original iterative update in \eqref{graprojsD} is equivalent to
\begin{equation}
\label{eq:dproj2}
\hspace{-1.6em}\begin{array}{ll}
\bby_{k+1}=\tProj_k[\bby_{k}-\epsilon(\tdbbX\bby_k+\bbu_k)],
\end{array}
\end{equation}
which can be viewed as the standard gradient projection update for the following dynamic constrained optimization problem
\begin{equation}\label{eq:prob2}
\min_{\bby\in \widetilde{\ccalQ}_k} \tdf_k(\bby):=\frac{1}{2}\|\bbP \bby + (\bbP^\T)^{-1} \bbu_k\|^2
\end{equation}
where $\bbP$ is obtained by the Cholesky factorization for the symmetric PD matrix $\tdbbX=\bbP^\T\bbP$. Correspondingly, $\bby_{k}^*$ is the optimizer of \eqref{eq:prob2}. Also, we denote
\[
C:=\min\limits_{k\geq0;\bby\in\R^N}\left\{\lmin{\nabla^2 \tdf_k(\bby)}\right\}=\lmin{\tdbbX}
\]
which is the smallest eigenvalue of $\tdbbX$. We say a differentiable function $f:\R^n\rightarrow\R$ is strongly convex with some positive constant $c$ if for any $x$ and $y$ we have $f(y)\geq f(x)+\langle\nabla f(x),y-x\rangle+\frac{c}{2}\|x\|^2$ where $\langle\cdot,\cdot\rangle$ is the inner product of two vectors; a differentiable function $f:\R^n\rightarrow\R$ is gradient Lipschitz continuous with some positive constant $m$ if for any $x$ and $y$ we have $f(y)\leq f(x)+\langle\nabla f(x),y-x\rangle+\frac{m}{2}\|x\|^2$. Note that, under this definition, $C$ also serves as the least strong convexity constant of $\tdf_k(\bby),\ \forall k$, while $M$ also serves as the greatest gradient Lipschitz continuity constant of $\tdf_k(\bby),\ \forall k$. A vector $v$ is called a subgradient of a convex function $f:\mathcal{X}\rightarrow\R\bigcup\{+\infty\}$ at point $x\in\mathcal{X}$ if $f(y)\geq f(x)+\langle v,y-x\rangle$ for any $y\in\mathcal{X}$. The set of all subgradients at $x$ is called the subdifferential at $x$. We use $\widetilde{\nabla}\tdf_k(\bby)$ and $\partial\tdf_k(\bby)$ to denote a subgradient and the subdifferential of the function $\tdf_k$ at $\bby$, respectively. These notations were also used in \cite{Bertsekas2011}. The subgradient used in the algorithm or analysis will be specified in the context, and our analysis will be based on the equivalent update and optimization problem in \eqref{eq:dproj2} and \eqref{eq:prob2}, respectively. The following lemma gives the first-order optimality condition of \eqref{eq:prob2} and an equivalent recursive relation of \eqref{eq:dproj2}.

{\re \begin{lemma}[First-Order Optimality Condition and Recursive Relation]
The instantaneous optimizer $\bby_k^*$ to the dynamic optimization problem \eqref{eq:prob2} and iterates $\bby_k$ satisfy the following conditions, for $k=0,1,\ldots$,
\begin{equation}\label{eq:opt_cond}
\tdbbX \bby_k^* + \bbu_k + \tdnabla g_k(\bby_k^*)=0
\end{equation}
and
\begin{align}
\label{eq:iterc}
\bby_{k+1}-\bby_k = -\epsilon \left[ \tdbbX(\bby_k-\bby_k^*) + \tdnabla g_k(\bby_{k+1}) - \tdnabla g_k(\bby_k^*) \right]
\end{align}
where
\[
g_k(\bby)=\left\{
            \begin{array}{ll}
              0, &\text{if $\bby\in\widetilde{\ccalQ}_k$},\\
              +\infty, &\text{if $\bby\notin\widetilde{\ccalQ}_k$}\\
            \end{array}
          \right.
\]
is the indicator function of the set $\widetilde{\ccalQ}_k$ at time $k$.
\end{lemma}}
\begin{IEEEproof}
We first replace the projection operation in \eqref{eq:dproj2} by a subgradient step featured by the indicator function $g_k(\cdot)$. By definition, the projection of any $\bbomega$ to $\widetilde{\ccalQ}_k$ equals to
\begin{equation}\label{eq:proj_def}
\tProj_k(\bbomega)=\arg\min_{\bbx} \epsilon g_k(\bbx) + \frac{1}{2}\|\bbx-\bbomega\|^2.
\end{equation}
The first-order optimality condition leads to $\epsilon \tdnabla g_k(\tProj_k(\bbomega)) + \tProj_k(\bbomega) - \bbomega = 0$. Thus by letting $\bbomega = \bby_k-\epsilon (\tdbbX \bby_k + \bbu_k)$ and $\tProj_k(\bbomega)=\bby_{k+1}$, we obtain
\begin{align}
\label{eq:firstor}
\bby_{k+1}=\bby_k-\epsilon [\tdbbX \bby_k + \bbu_k + \tdnabla g_k(\bby_{k+1})].
\end{align}

Furthermore, by using the indicator function, \eqref{eq:prob2} is equivalent to
\begin{align}\label{eq:eq-ind}
\bby^*_{k}=\arg\min_\bby \frac{1}{2} \|\bbP \bby + (\bbP^\T)^{-1} \bbu_k\|^2 + g_k(\bby).
\end{align}
From the first-order optimality condition of \eqref{eq:eq-ind}, we have $\tdbbX \bby_k^* + \bbu_k + \tdnabla g_k(\bby_k^*)=0$. This along with \eqref{eq:firstor} proves the recursive relation \eqref{eq:iterc}.
\end{IEEEproof}

{\re Note that the subgradient $\widetilde{\nabla} g_k(\bby_{k+1})$ used in \eqref{eq:firstor} is well-defined because (i) $\bby^{k+1}\in\widetilde{\ccalQ}_k$, (ii) $g_k(\cdot)$ is continuous over $\widetilde{\ccalQ}_k$, and (iii) the minimum in \eqref{eq:proj_def} is uniquely attained since $\|\bbx-\bbomega\|^2$ is real-valued, strictly convex, and coercive. Using the aforementioned notation, our analysis coincides with those earlier results on nonsmooth optimization; see e.g., similar notions and analysis schemes have appeared in \cite{Bertsekas2011,Shi2015,Davis2014} and references therein. Our main result is as follows.}

{\re \begin{theorem}[BIBO Stability with Geometric Decaying]\label{them:conv} Under Assumption \ref{ass:B-Drift}, for any step-size choice
\[
\epsilon\in\left(0,\frac{2}{C+M}\right],
\]
the expectation of the weighted tracking error between the decentralized control update $\bbq_k$ of \eqref{graprojsD} and the instantaneous optimal solution $\bbq_k^*$ can be bounded by
\begin{equation}
\hspace{-2em}\begin{array}{ll}\label{conv}
&\EE \|\bbq_{k}-\bbq_{k}^*\|_{\bbD^{-1}}^2
\leq \rho^k \EE \|\bbq_{0}-\bbq_{0}^*\|_{\bbD^{-1}}^2+\frac{1-\rho^k}{1-\rho}\Theta, ~\forall k
\end{array}
\end{equation}
where the geometric rate $\rho\in(0,1)$ and $\Theta$ is a bounded positive constant gap.
\end{theorem}}
\begin{IEEEproof}
By the smoothness and convexity of $\tdf_k$, it follows that \cite{Nesterov2013} (see \eqref{eq:M} for the definition of $M$)
\begin{equation}\label{eq:final_proof1}
\begin{array}{rcl}
&    &\frac{CM}{C+M}\|\bby_k-\bby_k^*\|^2 + \frac{1}{C+M}\|\tdbbX(\bby_k-\bby_k^*)\|^2\\
&\leq&\langle\bby_k-\bby_k^*,\tdbbX(\bby_k-\bby_k^*)\rangle.
\end{array}
\end{equation}
By applying the basic inequality
\[
2\langle \sqrt{\beta}\bba,\frac{1}{\sqrt{\beta}}\bbb\rangle\leq a\|\bba\|^2+\frac{1}{a}\|\bbb\|^2
\]
which holds for any $\beta>0$ and any real vectors $\bba$ and $\bbb$ of the same dimension, the right-hand-side of \eqref{eq:final_proof1} can be upper bounded by
\begin{equation}\label{eq:final_proof2}
\begin{array}{rcl}
& &\langle\bby_k-\bby_{k+1}+\bby_{k+1}-\bby_k^*,\tdbbX(\bby_k-\bby_k^*)\rangle\\ &\leq&\frac{C+M}{4}\|\bby_k-\bby_{k+1}\|^2 + \frac{1}{C+M}\|\tdbbX(\bby_k-\bby_k^*)\|^2 \\
& &+ \langle\bby_{k+1}-\bby_k^*,\tdbbX(\bby_k-\bby_k^*)\rangle.
\end{array}
\end{equation}
Substituting \eqref{eq:final_proof2} into \eqref{eq:final_proof1} leads to
\begin{equation}\label{eq:final_proof3}
\begin{array}{rcl}
&    &\frac{CM}{C+M}\|\bby_k-\bby_k^*\|^2\\
&\leq&\frac{C+M}{4}\|\bby_k-\bby_{k+1}\|^2 + \langle\bby_{k+1}-\bby_k^*,\tdbbX(\bby_k-\bby_k^*)\rangle.
\end{array}
\end{equation}
Since the indicator function $g_{k}(\cdot)$ is convex due to the fact that $\widetilde{\ccalQ}_k$ is a convex set in our settings, its subgradient $\widetilde\nabla g_k(\cdot)$ (subdifferential $\partial g_k(\cdot)$) is a (set-valued) monotone mapping (this can also be obtained from the subgradient inequality \cite{Davis2014}), i.e.,
\begin{equation}\label{eq:final_proof4}
\langle\bby_{k+1}-\bby_k^*,\tdnabla g_k(\bby_{k+1})-\tdnabla g_k(\bby_k^*)\rangle \geq 0
\end{equation}
Combining \eqref{eq:final_proof3} and \eqref{eq:final_proof4} we have
\begin{equation}\label{eq:final_proof5}
\begin{array}{rcl}
&    &\frac{C+M}{4}\|\bby_k-\bby_{k+1}\|^2 \\
&    &+\langle\bby_{k+1}-\bby_k^*,\tdbbX(\bby_k-\bby_k^*)+\tdnabla g_k(\bby_{k+1})-\tdnabla g_k(\bby_k^*)\rangle\\
&\geq&\frac{CM}{C+M}\|\bby_k-\bby_k^*\|^2.
\end{array}
\end{equation}
Substituting \eqref{eq:iterc} into \eqref{eq:final_proof5} for $\tdbbX(\bby_k-\bby_k^*) + \tdnabla g_k(\bby_{k+1}) - \tdnabla g_k(\bby_k^*)$ leads to
\[
\begin{array}{rcl}
&    &\frac{C+M}{4}\|\bby_k-\bby_{k+1}\|^2 + \frac{1}{\epsilon}\langle\bby_{k+1}-\bby_k^*,\bby_k-\bby_{k+1}\rangle\\
&\geq&\frac{CM}{C+M}\|\bby_k-\bby_k^*\|^2.
\end{array}
\]
Using the equality $\langle \bby_k^*-\bby_{k+1},\bby_{k+1}-\bby_k \rangle = \|\bby_k^*-\bby_k\|^2 - \|\bby_k^*-\bby_{k+1}\|^2 - \|\bby_{k+1}-\bby_k\|^2$ to expand the inner product, we have
\begin{equation}\label{eq:final_proof7}
\begin{array}{rcl}
\|\bby_{k+1}-\bby_k^*\|^2
&\leq& \left( 1-\frac{2 \epsilon CM}{C+M} \right) \|\bby_k-\bby_k^*\|^2 \\
&    &+ \left(\frac{\epsilon M+\epsilon L}{2} -1 \right)\|\bby_k-\bby_{k+1}\|^2.
\end{array}
\end{equation}
By choosing $\epsilon \leq \frac{2}{C+M}$ to ensure the second term on the right-hand-side of \eqref{eq:final_proof7} being nonnegative, the inequality \eqref{eq:final_proof7} can be further relaxed to
\begin{align}\label{eq:contraction_before}
\|\bby_{k+1}-\bby_k^*\|^2 \leq \left( 1-\frac{2 \epsilon CM}{C+M} \right) \|\bby_k-\bby_k^*\|^2.
\end{align}
By applying another basic inequality
\[
\|\bba^\prime+\bbb^\prime\|^2\leq(1+\beta^\prime)\|\bba^\prime\|^2+(1+\frac{1}{\beta^\prime})\|\bbb^\prime\|^2
\]
which holds for any $\beta^\prime>0$ and any real vectors $\bba^\prime$ and $\bbb^\prime$ of the same dimension, we have
\begin{equation}\label{eq:contraction_temp}
\hspace{-1em}\begin{array}{rcl}
&    &\|\bby_{k+1}-\bby_k^* + \bby_k^*-\bby_{k+1}^*\|^2 \\
&\leq&(1+\beta^\prime) \|\bby_{k+1}-\bby_k^*\|^2 + (1+\frac{1}{\beta^\prime})\|\bby_k^*-\bby_{k+1}^*\|^2 \\
&\leq&\rho \|\bby_k-\bby_k^*\|^2 + (1+\frac{1}{\beta^\prime})\|\bby_{k}^*-\bby_{k+1}^*\|^2
\end{array}
\end{equation}
where $\rho:=(1+\beta^\prime)( 1-\frac{2 \epsilon CM}{C+M})$ while the second inequality comes from \eqref{eq:contraction_before}. Let us denote $\Theta:=(1+\frac{1}{\beta^\prime})B_2$, and thus taking expectation on both sides of \eqref{eq:contraction_temp} gives us [cf. Assumption \ref{ass:B-Drift}]
\begin{equation}\label{eq:contraction_final}
\EE\|\bby_{k+1}-\bby_{k+1}^*\|^2 \leq \rho \EE\|\bby_k-\bby_k^*\|^2 + \Theta.
\end{equation}
Applying recursive induction on \eqref{eq:contraction_final}, we eventually obtain
\[
\EE \|\bby_{k+1}-\bby_{k+1}^*\|^2
\leq \rho^{k+1} \EE \|\bby_{0}-\bby_0^*\|^2+\frac{1-\rho^{k+1}}{1-\rho}\Theta,
\]
which recovers \eqref{conv} by the definition of $\bby_{k+1}$ and $\bby_{k+1}^*$.
It shows that as long as $\rho\in(0,1)$ and $\Theta\in[0,\infty)$, $\EE \|\bby_{k+1}-\bby_{k+1}^*\|^2$ is bounded for all $k$. Note that the choice of $\beta^\prime$ can be arbitrary close to $0$. Hence $\rho$ can always achieve a value that is less than $1$ as long as $\epsilon>0$ and $C>0$ (a simple choice to demonstrate this is $\beta^\prime=\epsilon CM/(C+M-2\epsilon CM)$). Finally, we conclude that the step-size condition is $0<\epsilon\leq\frac{2}{C+M}$.
\end{IEEEproof}

Theorem \ref{them:conv} establishes that the tracking error of the decentralized control update \eqref{graprojsD} under dynamical settings exponentially decreases until a constant error bound is reached. Moreover, the  AR(1) process assumed to  model the $\bbarbbv_k$ series can be potentially extended to a general stochastic process that has bounded iterative changes. This is because the constant $\Theta$ in \eqref{conv} is bounded as long as  the condition in \eqref{eq:var_dif_v} holds. To extract a more specific result, let us choose $\beta^\prime=\epsilon CM/(C+M-2\epsilon CM)$. In this case, the steady-state ($k\rightarrow\infty$) error is explicitly bounded by
\begin{equation}\label{eq:steady-state}
\begin{array}{rcl}
& &\lim_{k\rightarrow\infty}\frac{1-\rho^k}{1-\rho}\Theta\\
&=&\frac{1}{1-(1+\beta^\prime)(1-\frac{2\epsilon CM}{C+M})}(1+\frac{1}{\beta^\prime})B_2\\
&=&\frac{(C+M)(C+M-\epsilon CM)}{(\epsilon CM)^2}B_2.
\end{array}
\end{equation}
This error depends on system parameters $C$, $M$, step-size $\epsilon$, and the constant $B_2$ which bounds the successive difference of the instantaneous optimal solutions. It can be seen that the larger the step-size is, the smaller the steady-state error bound is. Letting the step-size be $\epsilon=\frac{2}{C+M}$ (best achievable) further yields that the steady-state error does not exceed $\frac{(C+M)^2(C^2+M^2) B_2}{4C^2M^2}$. To sum up, under the settings of both dynamic objective and dynamic constraint, the stable step-size is slightly smaller than the one in the static case (no optimality drift) but under both situations, the achievable step-sizes are on the same order $O\left(\frac{1}{M}\right)$ because $\frac{2}{M}\geq\frac{2}{C+M}\geq\frac{2}{M+M}=\frac{1}{M}$.
\begin{remark}[Time-Invariant Box Constraints] For the special case that the box constraints are time-invariant (only objective is time-varying), we can show that the stepsize choice to achieve stability is the same to the static case of $\epsilon\in(0,2/M)$ \cite{liu_ssp16}. This way, the same choice holds for static, dynamic, or asynchronous scenarios. Constant limits on reactive power are the case for photovoltaic inverters if the solar irradiance stays the same during e.g., night time and no-cloud scenarios.
\end{remark}

\section{Numerical Tests}\label{sec:num}

We investigate the performance of the decentralized voltage control scheme under the settings of asynchronous update and dynamically time-varying network operating conditions. The desired voltage magnitude $\mu_j$ is chosen to be $1$ at every bus $j$. Each bus is assumed to have a certain number of PV panels installed, and thus it is able to control its reactive power via advanced inverter design. All numerical tests are performed in MathWorks\textsuperscript{\textregistered} MATLAB 2014a software.

A single-phase radial power distribution network consisting of 21 buses is first used to test the algorithm. This network is equivalent to the system in Fig. \ref{fig:radial} for $N=20$. The impedance of each line segment is set to be  $(0.233+j0.366)\Omega$. Hence, the linearized flow equations in  \eqref{ldf} is only an approximate model. The limits of reactive power resources at every bus is chosen to be $[-100,~100]$kVA. More realistic test using a 123-bus multi-phase network will be presented later on.

\vspace*{3mm}
\noindent\textbf{Test Case 1:}
The impact of asynchronous updates across different buses is first considered under a constant nominal voltage $\bbarbbv_k$. The maximum update delay is set to be $K=50$. For the 21-bus network, the theoretical upper bound on the step-size is $\epsilon<2/M=0.0062$ following Theorem \ref{prop:asyn}. Hence, we set the step-size to be $\epsilon = 1/M = 0.0031$. To model the level of asynchronous updates among multiple buses, we introduce a duty cycle parameter $\eta\in(0,100\%]$.  For a cycle of total $\frac{K}{2}$ time slots, we randomly pick $\lceil\eta\times\frac{K}{2}\rceil$ number of slots for bus $j$ to implement its voltage control update. Hence, the maximum update delay among any two nodes is no more than $K$. In addition, the larger $\eta$ is, the more frequently every bus performs an update, and the smaller the effective update delay would be. In particular, the setting of $\eta=100\%$ provides the benchmark performance of the synchronous scenario where each bus updates at every time slot.
\begin{figure}[tb]
\centering
\includegraphics[width=0.9\linewidth]{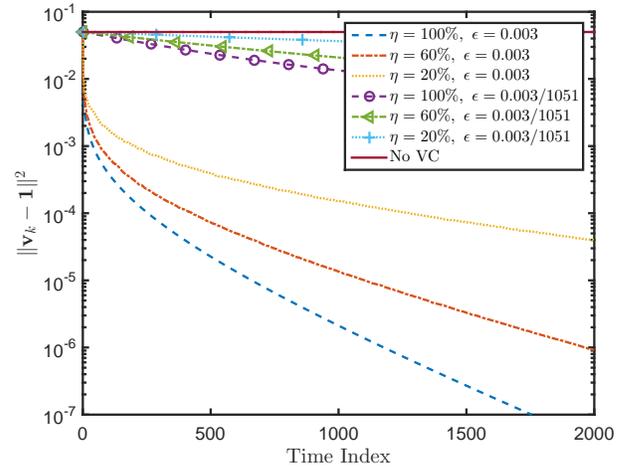}
\caption{Iterative voltage mismatch error performance for the asynchronous decentralized voltage control scheme under various choices of duty cycle $\eta$ and step-size $\epsilon$.}
\label{eps:traversing_eta}
\vspace{0em} \end{figure}
Fig. \ref{eps:traversing_eta} plots the iterative voltage mismatch error performance for the decentralized voltage control design  under different $\eta$ values and choices of step-size. The case of no voltage control is also plotted with the corresponding error staying constant. Using the classical convergence conditions for asynchronous GP updates in \eqref{eq:old_step_size},  the step-size should be chosen as $\epsilon=0.0031/1051$ with $N=20$ and $K=50$. As shown in Fig. \ref{eps:traversing_eta}, this choice of step-size is too conservative. Thus, the resultant convergence speed is much slower than that of the choice $\epsilon = 0.0031$ following Theorem \ref{prop:asyn}. This demonstrates that our theoretical results for asynchronous GP updates are more competitive for the specific decentralized voltage control application here.
\begin{figure}[tb]
\centering
\includegraphics[width=0.9\linewidth]{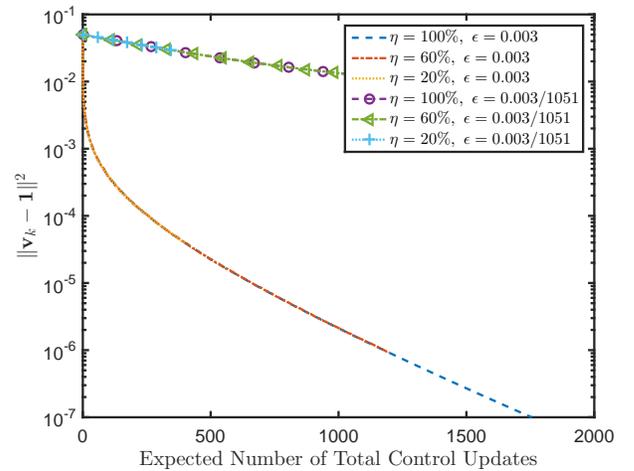}
\caption{Voltage mismatch error versus the total number of updates across the network for the asynchronous voltage control scheme under various choices of duty cycle $\eta$ and step-size $\epsilon$.}
\label{eps:traversing_eta_grad}
\vspace{0em} \end{figure}
Moreover, it is observed that the convergence accuracy would depend on the total number of updates for the whole network. Because of the asynchronous update settings, the expected number of control updates across the network for a cycle of $\frac{K}{2}$ iterations equals to $\lceil\eta\times\frac{K}{2}\rceil \times N$. Fig. \ref{eps:traversing_eta_grad} illustrates the voltage mismatch error performance versus the expected number of total control updates. Interestingly, the convergence speed in this plot is the same for the same step-size $\epsilon$ value, regardless of the asynchronous metric $\eta$. Hence, the average update rate across all the buses determines the performance of the asynchronous decentralized voltage control scheme.

\vspace*{3mm}
\noindent\textbf{Test Case 2:}
\begin{figure}[tb]
\centering
\includegraphics[width=0.9\linewidth]{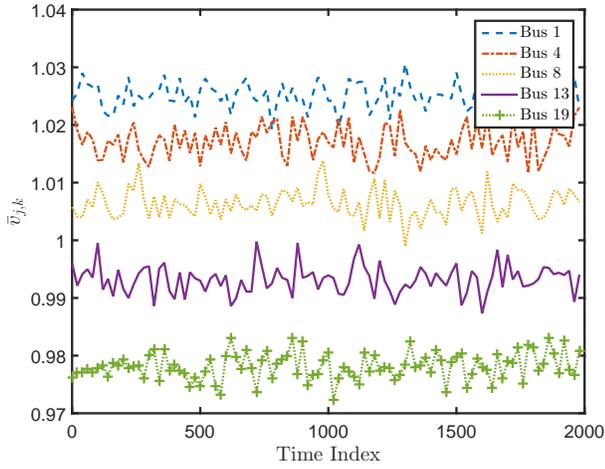}
\caption{The nominal voltage series $\{\bbarbbv_k\}$ at selected buses generated by the AR(1) model.}
\label{eps:voltpro}
\vspace{0em} \end{figure}
To verify our results on dynamic voltage control, we generate the nominal voltage series $\{\bbarbbv_k\}$ using the AR(1) model in \eqref{v_ar2}. {\ml Neglecting the effects of voltage regulators,} it is well known that the voltage magnitude in power networks tends to decrease monotonically away from the root node, i.e., bus 0. Hence,  we set the mean voltage at bus $j$ to be $\bbarc_j/(1-\alpha) = 1.025-\frac{0.05}{19}(j-1)$ to follow this decreasing voltage rule. 
Fig. \ref{eps:voltpro} plots the nominal voltage sequence at selected network locations for the choice of $\alpha = 0.1$ and noise variance $\sigma^2=6\times10^{-6}$. This choice of the forgetting factor $\alpha$ value leads very fast dynamics in the nominal voltage.

\subsubsection{The step-size $\epsilon$} \label{subsec:eps}
\begin{figure}[tb]
\centering
\includegraphics[width=0.9\linewidth]{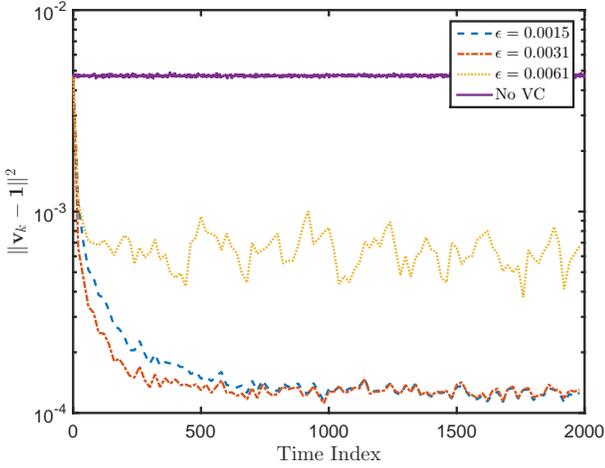}
\caption{Iterative voltage mismatch error performance averaged over 30 random realizations for the voltage control scheme with different $\epsilon$ values.}
\label{eps:traversing_epsilon}
\vspace{0em} \end{figure}
\begin{figure}[tb]
\centering
\includegraphics[width=0.9\linewidth]{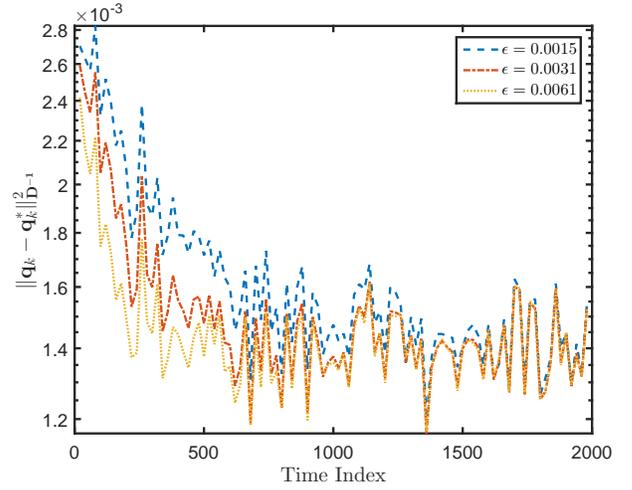}
\caption{Iterative tracking error averaged over 30 random realizations for the voltage control design with different $\epsilon$ values.}
\label{eps:traversing_epsilon_q}
\vspace{0em} \end{figure}
Fig. \ref{eps:traversing_epsilon} plots the iterative voltage mismatch error performance using different choices of $\epsilon$, while Fig. \ref{eps:traversing_epsilon_q} plots the weighted tracking error between the iterate $\bbq_k$ and the corresponding instantaneous optimal $\bbq_k^*$. Both plots are averaged over 30 random realizations of the nominal voltage series to approximate the expected values.
The maximum value $\epsilon=0.0061$ is chosen according to the bound $2/(C+M)$ in  Theorem \ref{them:conv}.
As shown more clearly in Fig. \ref{eps:traversing_epsilon_q}, a larger step-size $\epsilon$ leads to slightly faster convergence of the tracking error. However, the steady-state voltage mismatch error is higher for the largest $\epsilon$ as shown in Fig. \ref{eps:traversing_epsilon}. This observation coincides with the analytical results of Theorem \ref{them:conv}. The convergence geometric rate $\rho$ depends on an appropriate choice of $\epsilon$, while the steady-state error related constant $\Theta$ tends to increase with a larger $\epsilon$ choice. Thus, the choice of $\epsilon$  would be able to trade the steady-state tracking error off the convergence speed. Under this trade-off, the optimal selection of $\epsilon$ would also depend on the dynamics of the AR(1) process. If the nominal voltage evolves very fast, it is preferred to have a large $\epsilon$ for a  better tracking performance. Otherwise, if the dynamics of the nominal voltage series has a large time constant, we can afford to have a small $\epsilon$ in order to achieve a better tracking error performance. By analyzing the bounds in Theorem \ref{them:conv}, it is possible to provide a general guideline on selecting a proper $\epsilon$ value  based on the dynamics of $\bbarbbv_k$.

\subsubsection{The forgetting factor $\alpha$}
\begin{figure}[tb]
\centering
\includegraphics[width=0.9\linewidth]{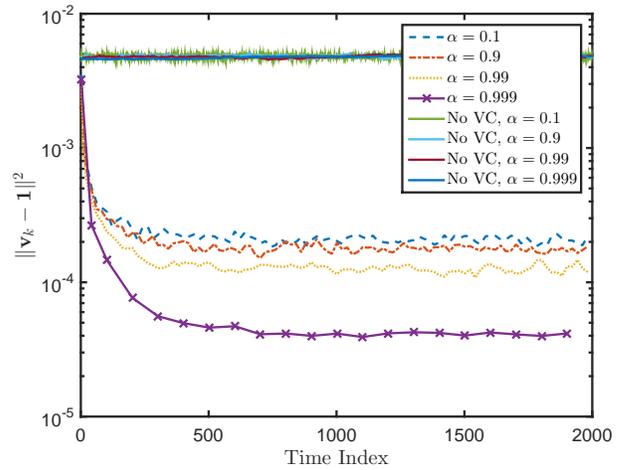}
\caption{Iterative voltage mismatch error performance averaged over 30 random realizations for the voltage control scheme with various values of forgetting factor $\alpha$.}
\label{eps:traversing_alpha}
\vspace{0em} \end{figure}
\begin{figure}[tb]
\centering
\includegraphics[width=0.9\linewidth]{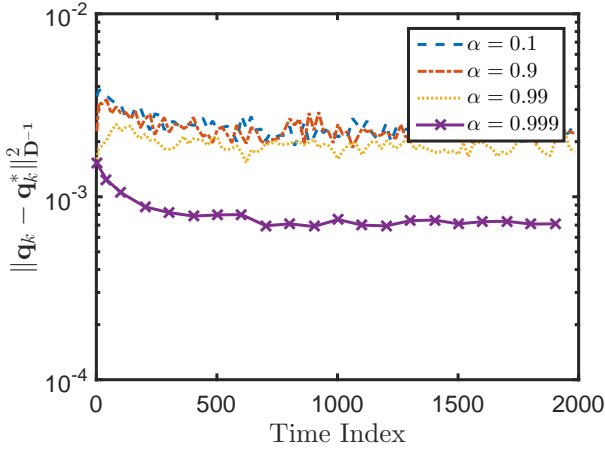}
\caption{Iterative tracking error averaged over 30 random realizations for the voltage control scheme with various values of forgetting factor $\alpha$.}
\label{eps:traversing_alpha_q}
\vspace{0em} \end{figure}
We have also varied the parameter $\alpha$ for the AR(1) process used to generate the nominal voltage series, with values ranging from 0.1 to 0.999. The step-size  $\epsilon$ is fixed at 0.0031. For comparison purposes, the variance of the nominal voltage at every bus is aligned to be the same for different $\alpha$ values by setting it  to be $\sigma^2/(1-\alpha^2) = 10^{-5}$. Hence, when $\alpha=0.999$ very closely approaches 1, the $\{\bbarbbv_k\}$ series would almost stays flat with minimal temporal variations. Accordingly, the consecutive voltage mismatch error bound $B_1$ in Prop. \ref{lemma:var_dif_v} decreases as $\alpha$ approaches its upper bound 1. Fig. \ref{eps:traversing_alpha} plots the iterative voltage mismatch error for various $\alpha$ values, while Fig. \ref{eps:traversing_alpha_q} again plots the corresponding weighted tracking error performance. These curves are also averaged over 30 random realizations.
{\ml The performance under either mismatch error metrics improves with a larger $\alpha$ value since the constant $B_1$ and thereby $B_2$ would decrease. Accordingly, this leads to a smaller $\Theta$ value and reduces the steady-state tracking error. This numerical result points out that $B_2$, which bounds the optimizer drift, could be related to the consecutive voltage difference $B_1$.}

\subsubsection{The noise variance $\sigma^2$}
\begin{figure}[tb]
\centering
\includegraphics[width=0.9\linewidth]{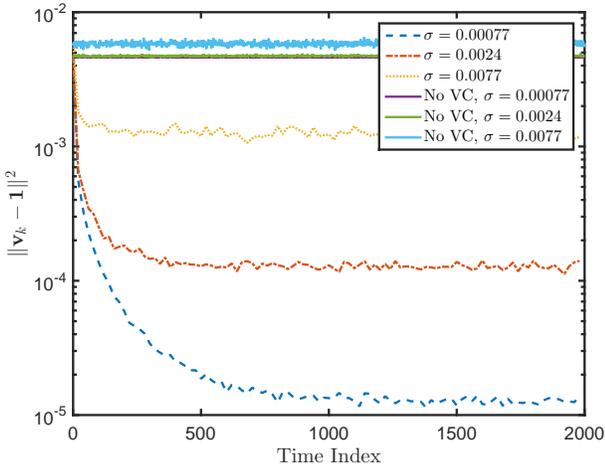}
\caption{Iterative voltage mismatch error performance averaged over 30 random realizations for the voltage control scheme with various $\sigma$ values.}
\label{eps:traversing_sigma}
\vspace{0em} \end{figure}
\begin{figure}[tb]
\centering
\includegraphics[width=0.9\linewidth]{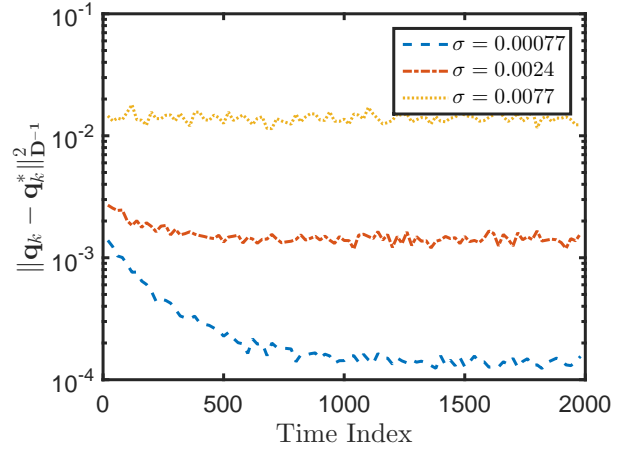}
\caption{Iterative tracking error averaged over 30 random realizations for the voltage control scheme with various $\sigma$ values.}
\label{eps:traversing_sigma_q}
\vspace{0em} \end{figure}
Similar test has been conducted with varying parameter $\sigma$ for the AR(1) process ranging from $7.7\times10^{-4}$ to $7.7\times10^{-3}$. With the same step-size $\epsilon=0.0031$ and parameter $\alpha = 0.1$, Fig. \ref{eps:traversing_sigma} and Fig. \ref{eps:traversing_sigma_q} plot the average voltage mismatch and weighted tracking error over 30 realizations. Similar to the observations under various $\alpha$ values, using a smaller $\sigma^2$ would decrease the steady-state error bounds since the constant $\Theta$ tends to be positively related to the parameter $\sigma^2$. As the parameter $\alpha$ is fixed, the variance of the nominal voltage at every bus decreases as the noise variance diminishes. Accordingly, the performance of no voltage control slightly improves with smaller noise variance.

All numerical results in this test case have verified our analytical  bounds on the tracking error performance. To sum up, the convergence speed depends on the choice of step-size $\epsilon$. Depending on the time constant of nominal voltage dynamics, the step-size needs to be properly chosen trading off the convergence speed and the steady-state error performance. The dynamics of nominal voltage series based on the AR(1) process parameters does not affect the convergence speed per se, yet more significantly related to the steady-state tracking error performance. Note that the analytical bounds of Theorem \ref{them:conv} are not tight, because of the scalar $\beta^\prime$ used to eliminate the cross-product terms from the squared sum norm. However, they are very effective to characterize the error performance of dynamic decentralized voltage control scheme while facilitating the selection of step-size.

\vspace*{3mm}
\noindent\textbf{Test Case 3:}
\begin{figure}[tb]
\centering
\includegraphics[width=0.9\linewidth]{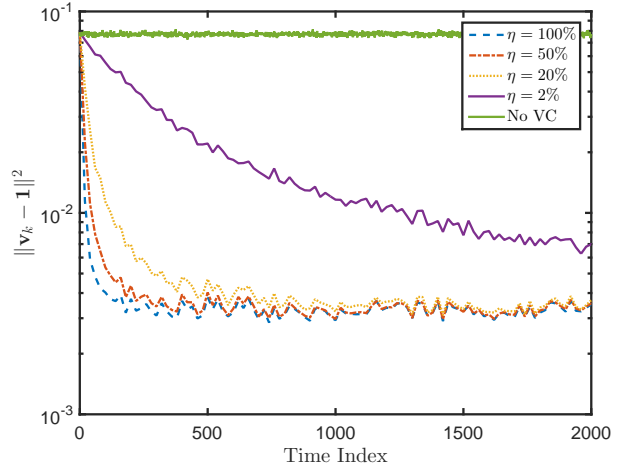}
\caption{IEEE 123-bus system iterative voltage mismatch error performance for the voltage control under both the asynchronous updates and the dynamic network operating conditions with different $\eta$ values.}
\label{eps:traversing_eta123}
\vspace{0em} \end{figure}
We have also tested the decentralized voltage control scheme using a realistic multi-phase distribution network, namely the IEEE 123-bus system model \cite{ieee123}. This test case incorporates both the asynchronous updates among multiple buses and dynamical nominal voltage, with similar settings as in the earlier two test cases. Moreover, the line segments of the 123-bus system are all lossy and involves inter-phase mutual couplings. Hence, this test provides an accurate representation of how the decentralized voltage control would perform in practice with uncertainties in the DER hardware and network operating conditions.

\begin{figure}[tb]
\centering
\includegraphics[width=0.9\linewidth]{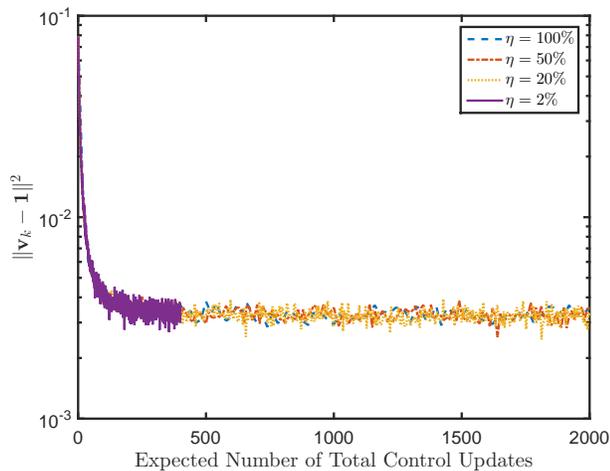}
\caption{Voltage mismatch error versus the total number of updates across the IEEE 123-bus system for the asynchronous voltage control scheme under dynamic operating conditions with various choices of duty cycle $\eta$.}
\label{eps:traversing_eta123_grad}
\vspace{0em} \end{figure}

Fig. \ref{eps:traversing_eta123} plots the iterative network voltage mismatch error under various choices of duty cycle parameters $\eta$. The step-size $\epsilon = 0.01$ has been chosen for every scenario to ensure stability. Different from the earlier two test cases, the control implementation has incorporated both the asynchronous updates and the dynamic voltage profile. Although we have not be able to derive the tracking error bounds under both sources of uncertainty, Fig. \ref{eps:traversing_eta123} demonstrates that its convergence speed results is similar to the solely asynchronous case as in Fig. \ref{eps:traversing_eta}, while the steady-state tracking error may have similar bounds as in the dynamic control analysis. In addition, Fig. \ref{eps:traversing_eta123_grad} illustrates the voltage mismatch error performance versus the expected number of total control updates. Similar to its single-phase counterpart, the convergence speed in this plot is analogous for a fixed step-size $\epsilon$ value, regardless of the asynchronous metric $\eta$. Thus, we are confident that the analysis of this paper can be integrated to a joint framework that characterizes the tracking error performance under both uncertain sources.

\section{Conclusions and Future Work} \label{sec:con}
This paper develops a decentralized dynamic optimization framework for analyzing the performance of a voltage control scheme based on gradient-projection (GP) methods for online system implementations. By constructing the linearized flow model for power distribution networks,  one can design a voltage control scheme by minimizing a surrogate voltage mismatch error using the GP iterations. Thanks to the physical power network coupling, this GP-based scheme boils down to a decentralized voltage control design where every bus can measure its local voltage to obtain the instantaneous gradient direction. Compared to earlier results for a static optimization scenario, we have significantly extended the analysis on convergence conditions and error performance to account for two dynamic scenarios: i) the nodes perform the decentralized update in an asynchronous fashion; and ii) the network operating point is dynamically changing. {\ml Assuming the nominal voltage evolves following an AR(1) process, the weighted tracking error can be bounded by an exponentially decreasing term plus a constant term that would depend on the successive difference of the transient optimal solution. Interestingly, the choice of step-size may need to be more conservative depending on the trade-off between the convergence speed and the steady-state tracking error for the dynamic control design.} Several numerical tests have been performed to demonstrate and validate our analytical results on the performance of the decentralized voltage control scheme under realistic dynamic scenarios using practical power network models.

Future work includes the development of an integrated framework for performance analysis under both dynamic scenarios simultaneously. We are also interested to investigate further on the impacts of vastly different time-scale among all network resources, which will help us to characterize the interactions with traditional voltage control devices of slower time-scales.

\bibliographystyle{IEEEtran}
\bibliography{IEEEabrv,admmref,stochvar,hzpub}
\end{document}